\documentstyle[aaspp4]{article}

\def\hst{{\sl HST \/}}
\def\iras{{\sl IRAS \/}}

\def\msun{M_{\odot}}

\def\um{\mu {\rm m}}

\def\jh{{\sl J--H\/}}
\def\jk{{\sl J--K\/}}
\def\hk{{\sl H--K\/}}
\def\k25{{\sl K--[25]\/}}
\def\flam{F_{\lambda}}
\makeatletter
\def\ale{\mathrel{\mathpalette\gl@align<}}
\def\age{\mathrel{\mathpalette\gl@align>}}
\def\gl@align#1#2{\lower.6ex\vbox{\baselineskip\z@skip\lineskip\z@
\ialign{$\m@th#1\hfil##\hfil$\crcr#2\crcr\sim\crcr}}}
\makeatletter
\begin{document}

\righthead{Axisymmetric PPN Reflection Nebulosities} 
\lefthead{Ueta, Meixner, \& Bobrowsky}

\title{An HST Snapshot Survey of Proto-Planetary Nebulae Candidates:
Two Types of Axisymmetric Reflection Nebulosities}

\author{Toshiya Ueta and Margaret Meixner}
\affil{Department of Astronomy, MC-221, 
University of Illinois at Urbana-Champaign, 
Urbana, IL  61801,
ueta@astro.uiuc.edu,
meixner@astro.uiuc.edu}

\author{Matthew Bobrowsky}
\affil{Orbital Sciences Corporation,
7500 Greenway Center Drive, \#700, 
Greenbelt, MD  20770,
mattbobrowsky@oscsystems.com}

\begin{abstract}
We report the results from an optical imaging survey of 
proto-planetary nebula candidates using the 
{\it Hubble Space Telescope} ({\sl HST}).
The goals of the survey were to image low surface brightness 
optical reflection nebulosities around proto-planetary 
nebulae and to investigate the distribution of the circumstellar 
dust, which scatters the star light from the central 
post-asymptotic giant branch star and creates the optical
reflection nebulosities.
We exploited the high resolving power and wide dynamic range 
of {\sl HST} and detected nebulosities in 21 of 27 sources.
The reduced and deconvolved images are presented 
along with photometric and geometric measurements.
All detected reflection nebulosities show elongation, and 
the nebula morphology bifurcates depending on the degree of
the central star obscuration.
The Star-Obvious Low-level-Elongated (SOLE) nebulae show 
a bright central star embedded in a faint, extended 
nebulosity, whereas the DUst-Prominent Longitudinally-EXtended
(DUPLEX) nebulae have remarkable bipolar structure with 
a completely or partially obscured central star.
The intrinsic axisymmetry of these proto-planetary nebula
reflection nebulosities demonstrates that the axisymmetry 
frequently found in planetary nebulae predates the
proto-planetary nebula phase, confirming previous independent 
results.
We suggest that axisymmetry in proto-planetary nebulae 
is created by an equatorially enhanced superwind at the end 
of the asymptotic giant branch phase.
We discuss that the apparent morphological dichotomy is caused by
a difference in the optical thickness of the circumstellar
dust/gas shell with a differing equator-to-pole density contrast.
Moreover, we show that SOLE and DUPLEX nebulae are 
physically distinct types of proto-planetary nebulae,
with a suggestion that higher mass progenitor
AGB stars are more likely to become DUPLEX proto-planetary 
nebulae.
\end{abstract}

\keywords{stars: AGB and post-AGB --- stars: mass loss  
--- planetary nebulae: general --- reflection nebulae} 

\section{Introduction}
Intermediate mass stars (initial main sequence mass of $0.8 - 8.0 \msun$) 
evolve through a transitional proto-planetary nebula (PPN) phase 
between the 
asymptotic giant branch (AGB) and planetary nebula (PN) phases (\cite{iben83}).
A PPN consists of a cool ($T_{\rm eff} \ale 10^4$ K) 
post-AGB stellar core and an extensive circumstellar shell of
gas and dust, which is the former stellar envelope ejected through 
wind mass loss\footnote{We reserve the word
``shell'' to refer to the circumstellar material that is physically
detached from the central star to avoid confusion with the word 
``envelope,'' with which we refer to the mantle of a star.}.
In the PPN phase, AGB mass loss is assumed to have ceased but 
photoionization of circumstellar matter is considered not to have 
been initiated (cf. \cite{kwok93}).
The PPN phase is still poorly studied because PPNe are statistically 
rare objects in the sky due to 
(1) their characteristically short time scale of evolution ($\sim 10^3$ 
years) with respect to a typical AGB evolutionary time scale ($\sim 10^6$ 
years) and
(2) technological constraints imposed by PPNe's 
small angular scale and the presence of dust grains.
Despite this rarity, approximately 100 PPN candidates have been identified
from optical and infrared studies (e.g., \cite{hrivnak89}, 
\cite{oudmaijer92}).

One of the most significant changes that occurs to stars during the 
transition is the emergence of axisymmetry.  
It is observationally established that most OH/IR stars, whose 
stellar cores are AGB stars (\cite{goldreich76}), show a high degree 
of spherical symmetry (cf. \cite{habing93}), while most ($80\%$) 
planetary nebulae display either bipolar or elliptical 
symmetry\footnote{We consider bipolar and elliptical morphologies
form mutually exclusive sets (see discussions below).}
(\cite{zuckerman86}).
Therefore, the departure from spherical symmetry must take place 
somewhere along the evolutionary sequence between the two phases.
The aspherical shaping of PNe has been qualitatively 
explained by the interacting stellar winds (ISW) model.
In this framework, the fast wind ($\age 10^3$ km/s) is expected to 
``snow plow'' slowly coasting ($\age 10$ km/s) circumstellar material 
which was ejected from the stellar envelope during the previous mass 
loss epoch (\cite{kwok82}).
Then, axisymmetry can be imposed by introducing the notion of the
equatorial density enhancement in the mass loss ejecta (\cite{kahn85}) 
and distinct morphological groups can be created by varying the degree 
of equatorial enhancement in red giant or AGB ejecta 
(e.g., \cite{balick87}, \cite{habing89}, \cite{mellema95}).
Although there has been a number of suggestions for the source of
equatorial density enhancement (which includes magnetic fields and
binary companions to name a few; e.g., \cite{soker98}, \cite{mm99}), 
there is no definite solution to the problem. 
Whichever the true scenario may be, a significant portion of 
the entire mass loss history is imprinted on the PPN circumstellar 
shell of gas and dust: 
the innermost edge defines the termination of mass loss
and the mass loss history can be traced back in time 
as one probes outer regions of the circumstellar shell.
Therefore, one can investigate when and how geometry of mass loss 
departs from spherical symmetry by sampling dust/gas distribution 
at various radial locations in a PPN circumstellar shell.
One must, however, employ techniques that are sensitive to neutral 
gas (molecular line emission in radio) and dust (thermal emission in 
infrared and scattering of star light in visible) since no 
photoionization has 
taken place in the PPN circumstellar shell.

In order to sample the most recent mass loss history one needs to probe
the innermost regions of a PPN circumstellar shell.
Previous ground-based investigations of PPN circumstellar shells
include 
mid-infrared imaging of the thermal emission from warm ($\sim 100$K) 
dust grains (e.g., \cite{meixner99}), 
optical imaging of the reflection nebulosity 
(e.g., \cite{hrivnak99b}),
and spectropolarimetry of the dust-scattered star light
(\cite{schmidt81}, \cite{johnson91}; \cite{trammell94},
hereafter TDG94).
All of these studies have shown the axisymmetric nature of the innermost
regions of PPN circumstellar shells.
Recent work on HST imaging of PPNe
(Egg Nebula, \cite{sahai98}; 
\iras 17150$-$3224, \cite{kwok98};
\iras 17441$-$2411, \cite{su98})
have displayed spectacular images of bipolar reflection nebulosity
but concentrated on a few PPNe that are associated with mostly or 
entirely obscured central stars.
Our \hst survey of PPNe covers a larger number of targets than any 
of the previous studies and, more importantly, includes PPNe candidates
that are associated with bright central stars.

Our main goal in this survey is to detect small and faint 
reflection nebulosities around PPN candidates at various evolutionary 
stages and to investigate 
if there exists any coherent morphological trend that will bridge gaps 
between the circumstellar shell morphologies in the AGB and PN 
phases.
In this paper, we report results of our survey of optical
reflection nebulosities around 27 PPN candidates, in which the
high resolving power and wide dynamic range of \hst are both 
exploited to the fullest extent.
In the following sections, procedures of observations and data reduction
are summarized (\S 2), results are presented (\S 3), and 
the physical nature
of morphological groups that we find and the subsequent implications
in the context of the PPN evolution are discussed (\S 4) 
with conclusions (\S 5).

\section{Observations}
\subsection{Modes of Observation}
The 27 PPN candidates were observed with the Wide Field and Planetary 
Camera 2 (WFPC2) on-board \hst between 1996 April and 1997 August 
(Program IDs 6364 and 6737) and were all acquired in the Planetary 
Camera (PC) chip (f/28.3, $0\arcsec.0455$ pixel$^{-1}$).
Observed coordinates for each object are listed in Tables 1, 2, and 3.

To obtain high resolution images, each source was observed
with a two- or three-point linear dithering pattern (the telescope is 
linearly shifted with a non-integer multiple of pixels).
The dithering technique is now widely used in WFPC2 observations 
to fully exploit the high resolving power of \hst from undersampled WFPC2 
images.
In conjunction with the dithering, the objects were observed with a 
set of different exposure times to cover the 
wide dynamic range that we required in our images.
To image faint nebulosities with a sufficiently high signal-to-noise 
ratio, observations were made with intentionally long exposure times.
For example, \iras 22272$+$5435 was observed with 0.11, 1.2, 16, and 120
sec exposures.
As a consequence, pixels in the vicinity of the central star were often 
saturated or contaminated by anomalies (e.g., bleeding, ghosts).
The short exposure frames, which would be free of saturated pixels, 
were used to recover the loss of information in such pixels.
In order to maximize the coverage of the dynamic range, we used a 
gain of 15 electrons per Data Number (DN) in our observations.

We typically used two wide band filters, F555W 
and F814W (WFPC2 equivalent of the Johnson-Cousins bands, 
{\sl V}$_{\rm J}$ and {\sl I}$_{\rm C}$; e.g. \cite{holtzman95}),
for each source.
However, F450W (Wide B) filter was used when a source had previously 
been observed with F555W. 
For extremely bright sources,
we used medium and narrow band filters, F410M (Str\"omgren $v$), 
F547M (Str\"omgren $y$), and F469N (\ion{He}{2}),
to avoid saturation in the shortest exposure frames.
In total, approximately a dozen raw images were obtained for
each source and filter.

\subsection{Data Reduction}
We used IRAF/STSDAS routines in data reduction.
A standard \hst pipeline calibration was performed with the 
latest reference files available at the time of data reduction.  
Duplicate frames of dithered images were combined into a single
image by applying the variable-pixel linear reconstruction
algorithm (``drizzle'' package v1.2, \cite{fruchter97}),
which would interlace each pixel in multi-point dithered frames
according to the statistical significance of each pixel.
The drizzled images were subpixelized ($0\arcsec.0228$ pixel$^{-1}$) 
during the process and thus the high spatial resolution was 
recovered.
Cosmic-rays were removed by the drizzling algorithm in the 
three-point dithered frames, while they were eliminated manually 
by replacing the contaminated pixels with a median of the 
neighboring pixels in the two-point dithered frames.

In order to create non-saturated final source images, 
the saturated and unsaturated frames were combined 
by replacing saturated pixels with unsaturated ones from the 
shorter exposures, scaled by the exposure times.
We have thus successfully obtained high dynamic range images of 
reflection nebulosities whose outer perimeters often seem to be sky-limited.
Figures 1, 2, and 3 show the reduced images.
The resulting nebular signal-to-noise ratio 
ranges from $\ale 2$ for a faint nebula to $> 100$ for a bright 
nebula, while the star-to-nebula ratio, which quantifies the emission 
contrast between the central star and nebulosity
as a measure of the dynamic range, ranges from 
$18$ for a compact nebula up to $1.8 \times 10^{5}$ for 
an extended nebula.
Some faint nebulae are barely distinguishable from the
background sky, while some others are almost buried 
under the point spread function (PSF) of the central star itself.

\subsection{Image Enhancements}
The reduced images, particularly those with a bright central star, 
were affected by the WFPC2 diffraction patterns that consist of 
linear spikes and circular wings.
We employed image enhancement techniques to remove
unwanted effects of the WFPC2 point spread function (PSF)
so that the reflection nebulae would be more clearly seen.
One could remove the effects of the WFPC2 PSF by either
(1) subtracting a stellar PSF from a source image or
(2) deconvolving a source image with a stellar PSF.

A PSF subtraction can be done with either an observed PSF or a 
synthesized PSF.
Although observed PSFs were often preferred (\cite{krist97a}), 
we used synthesized PSFs because there existed no PSF with a high 
enough dynamic range in proper filters for our observations.
Model WFPC2 PSFs were generated by the code Tiny Tim (v4.4, 
\cite{krist97b}) for a given mode of observation, but 
the PSF had to be scaled to account for the drizzling.

Alternatively, a deconvolution technique can be used to eliminate
the PSF effects in the reduced source images.
We used the Richardson-Lucy (RL) algorithm and maximum entropy 
method and the former generally yielded better results than the latter.
This suggests that the most significant noise in our images was 
due to photon shot noise because iterative solutions yielded by
the RL algorithm are known to converge into the maximum 
likelihood solution in Poisson statistics (\cite{shepp82}).
In general, the contrast between the nebula and sky emission 
at the outer perimeter of a nebula was increased by a
factor of $\age 2$; however, there seemed to be little 
improvement in a few very extended nebulosities where the nebula
emission was comparable to the sky emission.
Unlike the case of PSF subtraction, a deconvolution technique could 
be applied to all our images.
One shortcoming in RL deconvolution is that the 
algorithm is known to amplify uncertainties and generate false depressions 
around pixels with unusually high DNs, and such ``holes'' indeed appeared
in the deconvolved images.
Therefore, we have to keep in mind that any interior structure in the 
deconvolved images should be regarded as suspect.

Overall, we found that the RL deconvolved images 
provided the best removal of the PSF effects among all the methods
we tried.  Thus, we present only the RL deconvolved 
images alongside the original reduced images in Figures 1, 2, and 3.

\subsection{Measurements}
Photometric and geometric quantities were measured from the reduced
source images.
To give the total specific flux density ($F_{\lambda}$
in ergs cm$^{-2}$ s$^{-1}$ \AA$^{-1}$), the reduced images 
were flux calibrated by adopting the \hst photometric calibration of 
SYNPHOT (v4.0, \cite{simon97}).
First, we defined a photometric aperture that was large enough 
to encircle the entire source as well as the diffraction features.
The total DN of a source was then determined by summing all DNs 
within the aperture.
The background emission per pixel was estimated by calculating the 
averaged sky DN inside a 10-pixel-wide annulus that encloses the 
aperture but was separated from the aperture by a buffer zone.
The background DN, which is the averaged sky DN multiplied by 
the number of pixels in the aperture, was subtracted from the total 
DN only when the averaged sky DN per pixel was greater than the root 
mean square of the sky DNs in the background annulus because otherwise 
the background subtraction would introduce an additional $1 \sigma$ 
uncertainty to the results.
We expect that an uncertainty due to the background subtraction
is rather insignificant because considerably larger DNs in the
emission core will dominate the total emission of the source
and thus photon shot noise will dominate.
The total source DN was then converted into $\flam$ and the WFPC2 
system magnitude (STMAG).

The extent of the nebulosities was estimated from the images
by defining the ``edge'' of a nebula to be the outermost recognizable 
structure, in which the emission level turned out to be  
$1 \sigma$ up to about $7 \sigma$ of the sky 
depending on the quality of the image.
The major and minor axes of the nebula were measured and 
the ellipticity of a nebula was derived by $e = 1-b/a$
(a and b are respectively major and minor axis lengths).
With the edge of a nebula being defined, we can also measure
the surface intensity (ergs cm$^{-2}$ s$^{-1}$ \AA$^{-1}$ sr$^{-1}$) 
at the peak and edge of the nebula, from which we can obtain
the star-to-nebula surface intensity ratio as a measure of the 
width of the dynamic range covered by the image.
When the central star is visible, the peak coincides with the 
location of the central star, but, when the central star is 
totally obscured, the peak is simply the local maximum in
the emission region.
All derived quantities are summarized in Tables 1, 2, and 3.

\section{Results}
\subsection{Images of Reflection Nebulosity}
Of 27 PPN candidates, 21 were found with fascinating reflection 
nebulosities around the central stars and six did not seem to be associated
with any nebulosity.
By inspection, one immediately realizes the following:
(1) all 21 nebulae show asphericity with varying degrees
and 
(2) there clearly exist two types of axisymmetry among those
aspherical nebulosities.
One type of nebulosity is characterized by its very low surface
brightness and multi-axis elongations which surround extremely
bright central stars.
The other type, however, is distinguished by the limb-brightened bipolar 
lobes with their partially or completely 
invisible central stars somewhere in the nebulae.
Because this apparent bifurcation is so astonishing in the morphology 
of reflection nebulosity, we categorize the two types according 
to the traits in appearance described above and refer to the former 
type as the Star-Obvious Low-level-Elongated (SOLE) nebulae
while the latter type as
the DUst-Prominent Longitudinally-EXtended (DUPLEX) nebulae.
Among 21 nebulosities, 11 and 10 are respectively found to be 
SOLE and DUPLEX nebulae in this classification.

One of the key differences between SOLE and DUPLEX sources
is the star-to-nebula surface intensity ratio (Table 1 and 2),
which is a useful quantity to estimate the size of the dynamic 
range required to observe SOLE nebulae.
For sources with the visible central star (all SOLE
and DUPLEX nebulae with the partially visible central star), 
the dynamic range varies from 18 (for the most compact SOLE
source, \iras 07430+1115) to $1.8 \times 10^{5}$ (for the
most extended SOLE source, \iras 19114+0002) with the 
average value of $1.3 \times 10^{4}$.
On the other hand, for sources with the obscured central star
(bipolar DUPLEX nebulae), the dynamic range is at most
330 (\iras 20028+3910, whose northern lobe is barely detected) 
with the average value of 55.
In recent subarcsecond optical imaging of PPNe by ground-based
telescopes, four of our sources
(\iras 18095+2704, \iras 19374+2359, \iras 20028+3910, and 
\iras 22574+6609) were observed to determine their 
morphology (\cite{hrivnak99b}).
However, they were unable to determine if \iras 18095+2704
(the only SOLE source among the four) is
extended partly due to the brightness of the star (${\sl V}=10.3$) 
despite the suggestion of its extension from the FWHM
of their {\sl V} image.
Because of the wide dynamic range we achieved ($\sim 1000$),
our \iras 18095+2704 images clearly show that it is indeed
an extended source.
Therefore, to detect and image faint, extended reflection
nebulosities around a bright central star, i.e., the SOLE nebula, 
a very wide dynamic range must be used.
Even the six orders of magnitude coverage barely 
detects the outermost structure in images of \iras 19114+0002 
and there may be even fainter, more extended nebulosity.

Although we classify the objects mainly on morphological grounds,
spectral energy distribution (SED) and two-color diagrams
provide supplemental information to determine the morphological 
class of an object.
The sources that are not considered to be associated with any 
nebulosities are referred to as stellar sources (see discussions below).
We adopt these terms to address each morphological type 
hereafter and will discuss the morphological dichotomy in detail
in the following subsections.
Figures 1(a) to 1(c), 2(a) to 2(c), and 3 show SOLE, DUPLEX, and 
stellar sources, respectively\footnote{All fully reduced 
and deconvolved images are also available in FITS format on the 
World Wide Web's NCSA Astronomy Digital Image Library 
(ADIL, at http://imagelib.ncsa.uiuc.edu/document/99.TU.01/).}.

\subsubsection{SOLE Nebulae}
The SOLE nebulae show the very bright central star embedded
in a very low surface brightness nebulosity (Fig. 1).
This type of reflection nebulae has been imaged for the first 
time by this survey: the central star is so bright that the 
object would always appear as a point source unless the observations 
are done with the high enough resolution (available with {\sl HST}) 
and/or wide enough dynamic range (available with our method of multiple
exposure times).
The very eye-catching trident-like structures emanating from 
the central star are WFPC2 linear diffraction spikes and are
not to be construed as real structure.
The reader is encouraged to compare the images of the SOLE nebulae
(Fig. 1) with those of the stellar sources (Fig. 3) to help
the eye differentiate real structures from remnants of the PSF 
artifacts such as the diffraction spikes and circular halos.
The morphology of the SOLE nebulae can be further subdivided
into groups:
a simple ellipse, multiple ellipses (more than one ellipse superposed
onto one another with differing orientations of the major axes),
an ellipse with embedded bipolar structure, and an ellipse with
concentric shells.

\iras 07134+1005, \iras 17436+5003, and \iras 20462+3416 all have
large ($\ale 4\arcsec$) and faint nebulosities.
The size of the optical reflection nebulosity in \iras 07134+1005 is the
largest among the $21 \um$ feature sources (e.g. \cite{hrivnak91a}) 
and is comparable to mid-infrared images (\cite{meixner97}).
The extended nature of \iras 17436+5003 was suspected from 
wings of $^{12}$CO (J=2-1) and $^{13}$CO (J=2-1) line profiles 
(\cite{bujarrabal92}).
\iras 20462+3416, a young PN which has already started 
showing low extinction characteristics (\cite{parthasarathy93}, 
\cite{smith94}), 
was observed to have experienced a brief period of an
enhanced mass loss between 1993 and 1995 (\cite{garcialario97b}).
\iras 02229+6208 and \iras 07430+1115 have smaller (2\arcsec~and 1\arcsec)
nebulae.
Both of these sources, along with \iras 05341+0852, have recently 
been observed by Hrivnak \& Kwok (1999), but they were unable to 
determine if the sources are extended due to poor seeing. 
\iras 04296+3429, \iras 05341+0852, and \iras 22272+5435 have 
two axes of elongation that are not perpendicular to each 
other.
Despite a circular nebula prediction because of its shape of the SED
(\cite{hrivnak91a}, \cite{hrivnak99b}), 
\iras 04296+3429 shows a complex double-elongation structure:
the secondary E-W elongation of \iras 04296+3429 is close 
to but not aligned with a diffraction spike and is likely to be real.
This is, however, consistent with a suggestion that the source 
is associated with axisymmetrically distributed, optically 
thin dust (TDG94).
\iras 05341+0852 shows a diffuse elongation in the NE-SW direction
and there seems to be a secondary elongation inside of and tilted 
about 20$^{\circ}$ counter-clockwise from the primary one.
\iras 22272+5435, whose axisymmetric nature was already 
seen by spectropolarimetry (TDG94), has a bright, 
large core with the four elliptical 
tips which create an almost amoeba-like appearance for the nebula.
The northern and southern elliptical tips are of equal brightness,
but the western tip is 2.6 times fainter than the eastern tip,
which is approximately 1.6 times brighter than the northern and
southern tips: this suggests that the E-W elongation is 
tilted (with the eastern lobe being closer to us) but
the N-S elongation is not tilted.

\iras 06530$-$0213 and \iras 18095+2704 have rather peculiar
structures.
In addition to the well-defined elongation, both sources display 
an inner structure, which seems to be a limb-brightened bipolar 
lobes.
\iras 18095+2704 shows a similar spectropolarimetric trend as 
seen in \iras 04296+3429 (TDG94), which may be related to a 
secondary jet-like structure that extends in the NE-SW direction.
However, its very bright central star ({\sl V}$_{\rm WFPC2}=10.30$) 
and 
poor seeing prevented Hrivnak et al.\ (1999) from resolving
its extension.
\iras 19114+0002 shows rich structure: there are 
at least four inner concentric shells (11\arcsec, 7\arcsec, 
4\arcsec, and 3\arcsec) with some protuberance and 
one very sharp elongation ($8\arcsec.5$) about 15$^{\circ}$ 
East from North.  
The $^{12}$CO (J=2-1) map shows very extended structure, which seems to
have been shifting its direction: $10\%$ contour points to
72$^{\circ}$ East from North ($37\arcsec$), while $50\%$ contour 
points to 45$^{\circ}$ East from North ($18\arcsec$).
The protuberance seen in our images may have emanated from the same 
rotating point of origin.
The axisymmetric nature has also been seen in polarimetric 
observations (TDG94).
Although the sharp elongation suggests a rather large inclination
angle, \iras 19114+0002 is believed to be close to a pole-on
orientation, which is evidenced by a hollow shell structure seen in
both mid-infrared (\cite{hawkins95}) and near-infrared polarimetric 
(\cite{kastner95}) imaging studies.

\subsubsection{DUPLEX Nebulae}
The DUPLEX nebulae are recognized either by their magnificent bipolar 
nebulosities or by rather well-defined limb-brightened bipolar lobes 
(Fig. 2).
They are usually outlined by a lower surface brightness halo.
These nebulae differ from the SOLE nebulae in appearance primarily
because their central stars are partially or completely
obscured.
The diffraction spikes are not usually an issue in the images of
DUPLEX nebulae as the central stars are obscured from the direct 
view.
The DUPLEX nebulae can also be further subdivided into
two groups depending on the presence or absence of the central star.

\iras 16342$-$3814, \iras 17150$-$3224, \iras 17441$-$2411, 
\iras 20028+3910, and \iras 22574+6609 show multiple emission peaks
without clear indications of the central star's whereabouts.
Among those, \iras 17150$-$3224 and \iras 17441$-$2411 are found
with comparable lobes both in size and brightness and
their lobes possess some inner structure that seems to be point symmetric.
There are also thin concentric arcs extending beyond the perimeters 
of the lobes.
The arcs appear to be created independently of the lobes because 
the arcs maintain the same emission level both in and out of the 
lobes and hence seem to be unaffected by the presence of the lobes.
The intervals between arcs have been estimated to be too short to 
be caused by the consecutive AGB thermal pulses (\cite{pacynski75}).
Our {\sl B} and {\sl I} band images of \iras 17150$-$3224 and 
\iras 17441$-$2411 confirm the findings by previous \hst wide 
{\sl V} band observations (\cite{kwok98}, \cite{su98}).
Weintraub et al.\ (1998) obtained ${\rm H}_{2}$ emission profiles 
from these sources and confirmed their orientations suggested from 
the \hst optical images.
They even also found evidence of an expanding torus in 
\iras 17150$-$3224, which was shown in a {\sl V--I\/}
image constructed from ground-based observations
(\cite{kwok96}).
\iras 16342$-$3814 and \iras 20028+3910 have unequal lobes in which
some inner structure is recognized in the primary lobe.
The bipolar nature of \iras 16342$-$3814, an extreme AGB star, was 
revealed in H$_{2}$O and OH maser observations (\cite{likkel88}),
and recent VLA observations of OH maser determined its
inclination angle to be about 40$^{\circ}$ (\cite{sahai99}).
\iras 20028+3910 has recently been reported to be extended
($2\arcsec.2 \times 2\arcsec.0$, \cite{hrivnak99b}), but
this only corresponds to the S lobe (which is about 25 times 
brighter than the N lobe) of this bi-lobal object.
Interestingly, the deconvolved images of \iras 20028+3910 show 
multiple-peaks within the S lobe.
\iras 22574+6609 is optically resolved for the first time and the 
images indicate the presence of more than two emission peaks,
confirming an earlier suggestion of elongation (\cite{hrivnak99}).
The emission level of the suspected third emission peak 
($0\arcsec.2$ north of the second peak) is almost the same as that of 
the background sky and thus its presence is inconclusive. 
This ``third'' peak may simply be a part of a clumpy second
peak, in which case the overall appearance of the source
resembles that of \iras 17441$-$2411.
Our {\sl V} band photometry ({\sl V}$_{\rm WFPC2}=21.24$) differs 
from that of Hrivnak \& Kwok (1991a; {\sl V}$_{\rm J}=24$).
This difference is significant enough to mention, even though
we are comparing magnitudes in slightly different systems.
Their lower magnitude suggests that it may have been affected by the 
unusually poor seeing (\cite{hrivnak91a}), or this star may have 
been experiencing a significant brightening.

\iras 08005$-$2356, \iras 17423$-$1755, and \iras 19374+2359
have the partially visible central star with limb-brightened
bipolar lobes which appear as a pair of horseshoe structures
facing each other along the bipolar axis.
Slijkhuis et al.\ (1991) observed an unusually broad 
H$\alpha$ line profile in \iras 08005$-$2356 and attributed 
it to a fairly extended emission region.
This interpretation is independently supported by
spectropolarimetry, which shows an abrupt position angle shift 
suggesting an optically thick dust torus and optically thin
reflection lobes (TDG94).
Both of the above views are confirmed by the bipolar shape clearly
seen in our images.
Its SE lobe is approximately 3 times 
brighter than the NW lobe, which suggests that the SE lobe 
is tilted towards us so that the central star becomes 
partially visible within the conical opening angle of the 
lobe.
\iras 17423$-$1755 displays fascinating point symmetric jet-like 
structures extending $17\arcsec$ in the whole stretch.
The NW lobe is more prominent (8 times brighter) than
the SE lobe, whose presence can be traced with the help of
the slightly visible, outer part of the limb.
This suggests that the NW lobe is inclined towards us,
again explaining the partial view of the central star.
The way the SE lobe is obscured strongly suggests the
presence of a dust torus between the lobes.
This interpretation agrees with a model in an earlier
multi-wavelengths study, in which fast, collimated jets 
punctured a detached shell causing a torus-like shell 
structure (\cite{bobrowsky95}).
Although less prominent, there are at least three knots in 
each of the point symmetric jets as seen previously
(\cite{bobrowsky95}, \cite{riera95}, \cite{borkowski97}).
A hydrodynamic simulation shows a diverging outflow being 
focused into a narrow jet and the point symmetric structure
can be explained by wobbling jets (\cite{borkowski97}).
\iras 19374+2359 was observed by Hrivnak et al.\ (1999b)
and a round extension ($2\arcsec.6$) is seen.
This corresponds to the outer halo in our images.
Although our images of \iras 19374+2359 have smaller 
signal-to-noise ratio compared with other images,
one can discern the star from the nebula in the northern
lobe, which is 3 times brighter than the southern
counterpart, suggesting that the northern lobe is pointing
towards us.

The two remaining DUPLEX sources, \iras 09452+1330 
({\sl IRC}+10216) and \iras 23321+6545, 
do not neatly fit into either of the above subdivisions.
\iras 09452+1330 is the best studied C-rich AGB star in the Galaxy.
The {\sl I} band image was previously published (\cite{skinner98})
and is included in this survey for the sake of completeness.
The aspherical appearance of \iras 09452+1330 has been interpreted as
a bipolar nebula whose southern lobe is pointed towards
us, being separated from smaller northern lobes by a dust lane,
and the bright point-like source in the southern lobe may
be the central star (\cite{skinner98}).
Thus, we classify \iras 09452+1330 as a DUPLEX source
because its reflection nebulosity appears quite similar to
that of DUPLEX PPNe.
We also observed the source with the wide B filter but did not 
detect anything.
The optical counterpart to \iras 23321+6545 is imaged for
the first time.
Its very small spatial extent suggests that this object is located 
relatively far away.
However, the fact that this distant source appears extended
alternatively indicates that the nebulosity has 
rather high surface brightness compared to the central star, 
which is a typical characteristic of DUPLEX nebulae.
If \iras 23321+6545 were a SOLE nebula, the PSF of the bright 
central star would have masked any structure of the fainter,
compact nebula and the source would have appeared
as a point source.
Therefore, \iras 23321+6545 must possess DUPLEX structure,
possibly the one with the partially visible central star.
 
\subsubsection{Stellar Sources}
Figure 3 shows the sources lacking clear indications of the
presence of a nebulosity.
\iras 04386+5722, \iras 20043+2653, and \iras 22142+5206
only have the diffraction features and no evidence of 
extended emission regions.
There is no deconvolved image displayed for both \iras 20043+2653
and \iras 22142+5206 because all of the frames were saturated and 
reconstruction of the non-saturated peaks was not possible.
Ghosts appear in the images of \iras 04386+5722 as the ``double 
dots'' about $2\arcsec$ east of the star: they are double because
each ghost appeared at different chip locations in the dithered 
frames.
\iras 05113+1347 does not seem to have any extended nebula,
however, the deconvolved images leave rather high residual DNs
within $0\arcsec.4$ of the central star where a Tiny Tim PSF is 
able to simulate PSF effects rather well (\cite{krist97a}).
Because of its small angular extent, it is inconclusive whether
this is real or not.
\iras 10158$-$2844 and \iras 15465+2818 are considered to 
have had little recent mass loss but are associated
with very diffuse, extended circumstellar dust shells 
(\cite{gillett86}, \cite{waters89}).
Our images are consistent with this picture of little
mass loss by showing no apparent nebulosity.
Because these stars are very bright (hence the use of a narrow
band filter), the diffraction features are more prominent in these 
images than in other images.

\subsection{Measured Quantities and Binary Companions}
Tables 1, 2, and 3 summarize the measured
quantities for SOLE, DUPLEX, and stellar sources, respectively.
The quantities are
the total specific flux densities 
($F_{\lambda}$ in ergs s$^{-1}$ cm$^{-2}$ \AA$^{-1}$), 
WFPC2 system magnitudes (STMAG),
peak intensities, star-to-nebula intensity ratios,
sizes, and ellipticity.
Table 2 is subdivided into two sections according to the 
visibility of the central star:
the peak intensity represents the intensity of the star
when the location of the central star in the nebula is certain
(top 4 objects) whereas it represents the local intensity maximum 
in the emission region when the central star is completely or 
partially obscured (the rest).
For images that show clear bipolar structure with halo, 
the size and ellipticity are measured for the entire halo as well 
as for each lobe.
We only list photometric quantities for stellar sources (Table3).

All {\sl V}$_{\rm WFPC2}$ are generally in good agreement 
with the previously published {\sl V}$_{\rm J}$ in the 
literature ($|\delta {\sl V}| \approx 0.18$ mag)
where the STMAG closely resembles the Johnson system,
except for two sources (\iras 16342$-$3814 and \iras 22574+6609).
Comparison between {\sl I}$_{\rm WFPC2}$ and available 
{\sl I}$_{\rm C}$ in the literature suggests that {\sl I}$_{\rm WFPC2}$
is generally $\sim 1.3$ dimmer than {\sl I}$_{\rm C}$.
This offset is due to the definition of STMAG system 
and the amount of offset is about equal to what is expected
by definition in SYNPHOT (\cite{simon97}).
Similarly, magnitudes obtained with other filters deviate
from the values in the literature due to the definition of 
the STMAG system.

Our {\sl V}$_{\rm WFPC2}$ of \iras 16342$-$3814 (15.64) 
agrees with
another measurement independently made by Sahai et al.\ (1999)
from the same image (15.7).
However, these values are significantly dimmer than the 
value reported in a PPN photometric survey (13.65; \cite{veen89},
hereafter VHG89).
A large photometric aperture used by them is suspected to have 
included nearby bright stars (\cite{sahai99}).
On the other hand, their near-infrared photometric values
({\sl J}=12.17, {\sl H}=10.75, {\sl K}=9.61, done in 1986; VHG89) 
are significantly dimmer than those of yet another PPN photometric 
survey ({\sl J}=9.29, {\sl H}=8.32, {\sl K}=7.71, done in 1993; 
\cite{garcialario97}).
Because Garc\'{\i}a-Lario et al.\ (1997) do not discuss 
their sources individually or give the observed coordinates,
we are unable to assess the cause of the discrepancy. 
It is very unlikely that the central star of a PPN 
becomes dimmer in {\sl V} and brighter in near-infrared.
\iras 22272+5435 is suspected to be a variable star
with an almost 1 mag variation (\cite{hrivnak91b}).  
Our measurement ({\sl V}$_{\rm WFPC2}=8.63$) is consistent with
its brightest magnitude.
There seems to be no sign of any other unreported variability in our 
sources.

The size of SOLE nebulae is the major and minor axis lengths of the
elliptical elongation whereas the size of DUPLEX nebulae is the extent of 
the halo.
Some SOLE nebulae have two axes of elongation. 
In such cases, we only list major axis lengths of the two elongations
but not the minor axis lengths, though the ellipticity are given for each.
The averaged ellipticity is rather high for both SOLE and DUPLEX 
nebulae (0.45 and 0.43, respectively).
Here, the ellipticity is even larger for SOLE nebulae.
This quantitatively confirms what we have seen in the images:
reflection nebulosities are unquestionably elongated in their
apparent shapes.

We can also search for signs of binary companions in the vicinity
of the sources, and there are several possible cases.
\iras 10158$-$2844 is seen with a star (WFPC2 \ion{He}{2} mag $=13.70$)
which is about $2\arcsec$ east of the source.
Although \iras 10158$-$2844 is known to form a binary system of an 
orbital period of near 434 days with either a low-mass main sequence 
star or a white dwarf (\cite{waelkens91}),
this nearby star does not seem to be the companion of the binary 
system because the separation is too large ($\approx 1500$AU).
Whether or not this nearby star is related to \iras 10158$-$2844
is also not certain.
\iras 19114+0002 has a nearby star seen about $4.5\arcsec$ north of the
source at the edge of the faintest nebulosity along one of the
saturation spikes.
However, the stars do not seem to be related due to rather large 
distance between the two.
\iras 19374+2359 has a nearby star ({\sl V}$_{\rm WFPC2}=20.39$) 
inside the south lobe about 1\arcsec~away 
from the central star, but it is likely to be a foreground or 
background star because of the low galactic latitude of the source 
($1^{\circ}$).
\iras 22142+5206 also has a star ({\sl V}$_{\rm WFPC2}=20.36$) 
about 2\arcsec~east of the source, but the nature of the nearby
star is not certain.
\iras 22142+5206 is now classified as a young stellar object 
embedded in a massive molecular cloud ($\sim 7300 \msun$; 
\cite{dobashi98}).
Their observations indicate that the source is associated with the
most massive CO outflow ($\sim 33 \msun$) reported
so far, and this may be because of the binarity of the source.

\section{Discussions}
\subsection{Axisymmetry: an Intrinsic Nature of PPNe}
We have detected optical reflection nebulosities in 21 sources out 
of 27 PPN candidates ($78\%$) and 
all of these 21 PPN reflection nebulosities exhibit some 
type of axisymmetry with the averaged ellipticity of 0.44.
Our direct imaging of reflection nebulosities confirms previously
published results in a spectropolarimetric survey of post-AGB 
stars, which has indirectly shown that 24 of 31 sources ($77\%$) 
are aspherical (TDG94).
As we have discussed in the introduction, previous work in 
the literature has revealed that most PPN candidates possess 
axisymmetric nebulosities,
and therefore, we conclude that the axisymmetry is an intrinsic
trait of the PPN reflection nebulosities.
Below, we discuss when this axisymmetry arises along the 
evolutionary track between the AGB and PN phases and
how the two different types of PPN may arise.

It is now generally accepted that the AGB phase is associated 
with two types of mass loss: an AGB wind ($\sim 10$ km/s) 
mass loss phase followed by a briefer but supposedly more violent 
superwind ($\sim 20$ km/s) mass loss phase (\cite{renzini81}).
Because the termination of a superwind is considered to be the end of 
the AGB phase and no significant stellar wind is expected until the 
initiation of a fast wind (\cite{kwok82}), the PPN axisymmetry
must arise just before the end of the AGB phase.
This interpretation is also supported by the fact that mass loss 
is spherically symmetric in the beginning of the AGB phase
(\cite{habing93})
and that some extreme AGB stars have already departed from 
spherically symmetric structure (e.g., \iras 16342$-$3814).
It is, therefore, very likely that a superwind is intrinsically
axisymmetric and that the onset of a superwind initiates
the morphological shift from spherical to axial symmetry 
in a PPN circumstellar dust/gas shell.

Based on the results of our PPN survey with the above inference,
we propose the following evolutionary scenario.
In the AGB wind phase, an AGB star loses its mass through a 
dust-driven AGB wind (\cite{salpeter74}, \cite{kwok75}, 
\cite{netzer93}) in a largely spherically symmetric manner, 
creating a spherically symmetric circumstellar AGB wind shell.
Towards the end of the AGB phase, some physical mechanism,
albeit still unknown, comes to play and starts generating an
equatorially density enhanced dust-driven wind, which we 
call an axisymmetric superwind.
The axisymmetric superwind dumps the envelope material of 
the central AGB star preferentially on the equatorial plane, 
and a superwind shell with a torus-like density enhancement
develops deep within the spherically symmetric AGB wind shell.
The equatorial density enhancement in the superwind shell is 
further strengthened as the star evolves.
At the end of the AGB phase, the superwind ceases and defines
the inner boundary of a detached circumstellar dust/gas shell,
which manifests itself as a mid- to far-infrared 
excess in the double-peaked SED of a PPN (\cite{kwok93}).

This two-phased AGB mass loss scenario can be employed to explain 
an apparent ``dust lane'' obscuring the central star.
In one-dimensional radiation transfer simulations, the superwind
shell may be treated as a somewhat {\it ad hoc} addition to an
otherwise spherically symmetric dust shell (e.g., \cite{su98}).
Meixner et al.\ (1997) have incorporated the two-phased
mass loss in fully two-dimensional radiation transfer 
calculations and their synthesized mid-infrared images and SEDs
agree with observations.
However, the scenario used in these calculations is still 
a first order approximation: 
the transition from a spherically symmetric AGB wind to an 
axisymmetric superwind is assumed to take place abruptly.
Instead, the transition is more likely to occur gradually 
because mass loss is essentially governed as a function of 
the fundamental stellar parameters, which do change gradually 
as the central star evolves if integrated over the course 
of the entire AGB phase (e.g., \cite{blocker95}).

Given that the details of the two-phased mass loss probably
depend on 
the fundamental stellar parameters, the degree of equatorial 
density enhancement in a superwind may also be dependent upon them,
and the physical environment will probably be distinct in each 
superwind shell.
Figure 4 schematically describes how this can affect the 
structure of PPN circumstellar shells and may cause the 
morphological bifurcation of PPN reflection nebulosities. 
In a SOLE PPN (top),
a marginal equatorial enhancement in the superwind shell 
can yield a dust torus that is optically thin (gray zone), 
and stellar photons can escape virtually in all directions (arrows).
Hence an observer is able to see the bright central star embedded 
in an elliptically elongated nebula (dashed perimeter).
In a DUPLEX PPN (bottom), on the other hand, a stronger equatorial 
enhancement in the superwind shell can result in
a optically thick dust torus (black zone),
and most photons are scattered off towards the biconical openings of the
torus along its axis of symmetry (arrows), thereby generating 
a bipolar, dumbbell-like nebulosity (dashed perimeter).
When the equatorial enhancement is exceptionally strong,
the dust torus can be so flattened that it assumes the form of 
a thin disk.
Therefore, we consider a disk to be an extremely 
equatorially enhanced torus.

\subsection{SOLE vs. DUPLEX: Physically Distinct Nebulae}
The distinct appearances between SOLE and DUPLEX nebulae
are characterized by the presence or absence of the central star 
and by the undisturbed elliptical or dumbbell-like outline of 
the nebulosity.
We now discuss how
the evident morphological dichotomy of the PPN reflection 
nebulosities can be caused by a physical difference in 
the circumstellar dust/gas shell
and not by an inclination angle effect alone.
More specifically, we propose that the optical morphology of
PPN candidates bifurcates because
the opacity in the circumstellar dust/gas shells varies 
due to differing degrees of equatorial density enhancement.
Spectropolarimetric survey results of TDG94 show that there 
are two types of polarization position angle shift (a gradual 
shift in \iras 04296+3429, a SOLE source and an abrupt shift in 
\iras 08005$-$2356, a DUPLEX source) and that the presence of 
the optically thick and thin, two-component obscuring agent 
is suspected in PPNe with an abrupt position angle shift.
This is consistent with the assumption of DUPLEX nebulae
being associated with optically thick dust grains.
In the following, we will present three additional pieces of 
evidence suggesting that the morphological dichotomy indeed 
corresponds to physically 
distinct nature of the circumstellar dust shell in PPN nebulae:
the mid-infrared morphologies, SEDs, and
{\sl IRAS}/near-infrared colors.

\subsubsection{Mid-Infrared Morphology of PPN Dust Shells}
Meixner et al.\ (1999) have recently observed 66 PPNe at mid-infrared 
wavelengths and directly imaged thermal dust emission regions
in the circumstellar dust shell.
The major discovery in the mid-infrared survey is the morphological 
bifurcation of dust emission regions: ``core/elliptical'' types have 
an extended low emission region surrounding a compact unresolved
core, which is attributed to an optically thick equatorial density 
enhancement, 
while ``toroidal'' types show two emission 
peaks, which are interpreted as limb-brightened peaks of an 
optically thin, equatorial density enhancement.
If we compare the mid-infrared and optical morphologies 
of PPNe, there is a one-to-one correspondence 
between the two morphologies as is shown in 
Table 4\footnote{We included four other PPN candidates (one 
SOLE source, \iras 21282+5050, and three
DUPLEX sources, \iras 04395+3601, Red Rectangle 
(\iras 06176$-$1036), and
Egg Nebula) whose mid-infrared and optical
morphologies have been identified.}.
It appears that toroidals and core/ellipticals are strongly 
correlated with SOLE and DUPLEX nebulae, respectively.
This correlation is consistent with the picture in which
a dust optical thickness difference causes the PPN morphological
bifurcation.
That is, PPN candidates that are optically thin at visible 
wavelengths (SOLE nebulae) are also optically thin at mid-infrared 
wavelengths (toroidals), while PPN candidates that are optically thick at 
visible wavelengths (DUPLEX nebulae) are also optically thick at 
mid-infrared wavelengths (core/ellipticals).

The ways in which the mid-infrared and optical images are spatially 
related in SOLE and DUPLEX nebulae differ, and hence, they also 
suggest a difference in dust shell optical thickness.
By direct comparison between the mid-infrared and optical images of 
SOLE sources, the optical and mid-infrared nebulae are found to be 
spatially coincident
and that the optical nebulosity is elongated perpendicularly with 
respect to the equatorial plane of the suspected dust torus, whose
orientation is indicated by the two mid-infrared emission peaks.
In Figure 5, for example, a resolved, deconvolved $11.8\um$ image 
of \iras 07134+1005 shows limb-brightened dust emission 
peaks which 
are oriented in the east-west direction (\cite{meixner97}),
while its optical (Str\"omgren {\it v}) nebula is extended in the 
north-south direction (top left).
A similar trend is seen in the composite image of \iras 17436+5003 
(top right; \cite{skinner94}) as well.
On the other hand, the mid-infrared and optical images of DUPLEX 
sources show elongation in the same direction, and the mid-infrared 
emission region is often completely embedded within the optical 
nebulosity.
In Figure 5, for example, an {\sl I} band image of \iras 17150$-$3224 
clearly displays the dust lane, which completely obscures the central 
star, between the two lobes and there is a very compact, unresolved
mid-infrared emission core over the location of the dust lane
(bottom left), 
and the unresolved dust emission core of \iras 16342$-$3814
in 9.8$\um$ is very compact with respect to the whole extent of 
the {\sl I} band reflection nebulosity (bottom right).

\subsubsection{Spectral Energy Distribution of PPNe}
One of the well-established characteristics of PPNe is that
their SEDs have a 
``double-peaked'' structure (cf. \cite{kwok93}).
The shortward and longward peaks in the wavelength spectrum 
respectively
correspond to the stellar and dust emission components.
The morphological bifurcation between the SOLE and DUPLEX 
nebulae also manifest itself as a distinction between the 
SED shapes of these sources.
A number of post-AGB stars have been classified into four classes 
based on the shape of the SED (VHG89):
\begin{enumerate}
\item[I.] Flat spectrum between 4 and 25$\um$ and a steep
fall-off to shorter wavelengths,
\item[II.] Maximum around 25$\um$ and a gradual fall-off to shorter
wavelengths,
\item[III.] Maximum around 25$\um$ and a steep fall-off
to a plateau roughly between 1 and 4$\um$ with a steep
fall-off at shorter wavelengths,
\item[IV.] Two distinct maxima; one around 25$\um$ and a
second between 1 and 2$\um$ ({\bf IVa}) or one around 25$\um$ and a
second $<1\um$ ({\bf IVb}).
\end{enumerate}
Here, we adopt the VHG classification scheme for the SEDs
of our sources. 
Because only five of our sources were studied in VHG89 with a 
partial coverage of the stellar component (i.e., $> 1\um$),
we compiled new SEDs of our sources by adding our photometric
measurements at optical wavelengths to the latest published 
data in the literature. 
Figures 6, 7, and 8 show updated SEDs for SOLE, DUPLEX, and stellar 
sources, respectively, with VHG class assignments indicated
in each frame. 
The SED shapes of SOLE and DUPLEX nebulae 
are indeed very distinct from each other, but are very similar within 
each morphological type of the nebulosity.
Stellar sources are of a mix of classes and are discussed
in \S 4.2.5.
Table 4 (column 5) summarizes a clear correlation between the
morphological and SED classes.

All SOLE nebulae have a double-peaked, class IV SED.
The prominent central stars in SOLE nebulae appear in their SED as 
the well-defined, unobscured optical/near-infrared peak
and the thermal emission from the circumstellar dust appears as an
almost equal flux peak.
There is a subdivision of the SED class among SOLE nebulae depending 
on the location of the optical/near-infrared peak.
The difference stems from the degree of reddening of the central
star due to its circumstellar dust/gas shell.
Class IVb PPNe tend to have physically larger circumstellar shells 
and hence their column densities are lower, less dereddening the 
central star.
Interestingly, mid-infrared images are
resolved for those of class IVb (\iras 07134+1005, \iras 17436+5003, 
and \iras 19114+0002), and this coincidence is consistent with
the picture of more extended, optically thin dust shells of
SOLE nebulae.
\iras 21282+5050 is a very young PN that we classify 
as a SOLE source based on morphology (\cite{khl93}, \cite{meixner97}), 
but its SED class appears to be III because the shorter wavelengths
light from its hot central star (Of7, \cite{cohen87}) is
more reddened by the dust than the star light from a typical PPN 
central star. 

DUPLEX SEDs are of class II (e.g., \iras 19374$+$2359) or of class 
III (e.g., \iras 17150$-$3224).
Both classes II and III are characterized by a prominent 
far-infrared peak (30 to 50$\um$) with an optical/near-infrared 
excess that represents the central star or its associated reflection
nebulosity.
The difference between class II and III SEDs is the presence of a 
rather large near-infrared excess in class II, 
which is commonly attributed to either an ongoing mass
loss episode or the presence of very compact circumstellar
dust shell (VHG89).
Interestingly, this SED class difference among DUPLEX sources 
corresponds to the visibility of the central star.
That is, the SED will be of class II when the central star 
is partially visible whereas it will be of class III when the central
star is completely obscured from the view.
\iras 09452+1330 is of class I but has a flat peak 
at $\ale 10\um$ caused by its lower $T_{\rm eff}$ ($\sim 2000$K),
which is consistent with its AGB stellar nature.
Its very sharp drop into the shortward wavelengths resulted in 
non-detection in our {\sl B} band image (Table 2). 
We classify \iras 08005$-$2356 of class II because of its
gradual fall-off in shorter wavelengths due to rather large 
optical/near-infrared flux.
Our classification of \iras 08005$-$2356 and \iras 17423$-$1755
being DUPLEX sources is well supported by the resemblance in the 
shapes of the two SED classes.

The differences in the SED shapes between SOLE and DUPLEX nebulae
can be explained in the context of our hypothesis.
In the case of a SOLE nebula, the circumstellar dust is optically thin 
and permits a clear, albeit reddened, view of the central star
with a modest amount of dust emission, which is proportional to the
column density of dust.
Hence, we see two distinct, comparable peaks of stellar and dust 
emission in its SED.
In the case of a DUPLEX nebula, on the other hand, the circumstellar dust
is optically so thick that almost all of the star light is absorbed by
the dust and is reradiated at mid- to far-infrared wavelengths;
only a few optical photons escape through the biconical openings of
the dust shell.
Therefore, we see a prominent dust emission peak accompanied by an
optical/near-infrared plateau in its SED.
In the framework of our hypothesis, an inclination angle effect
among DUPLEX sources manifests itself as the difference in 
their SED shapes.

\subsubsection{{\sl IRAS}/Near-Infrared Two-Color Diagrams}
To demonstrate the differences between SOLE and DUPLEX sources,
we use the \jk~vs. {\sl K--[25]} diagram, an {\sl IRAS\/}/near-infrared 
two-color diagram (Fig. 9).
Here, [25] is \iras flux at $25\um$ in magnitude converted by
$[25]=-2.5 \log(F_{\nu}/6.73)$ (\cite{iras}).
Because \k25~color relates the heights of stellar and dust peaks
whereas \jk~color describes the shape of the stellar component,
the \jk~vs. \k25 diagram introduces characteristics of detached
dust shells and incorporated the SED dichotomy into a diagram.
The robustness of the \jk~vs. \k25 diagram is evident in the
clear bifurcation between morphological groups.
In \jk~color, SOLE sources are bluer ($\ale 1.45$) than DUPLEX 
sources.
This bluer color is consistent with the presence of an optically 
thinner circumstellar shell along the line of sight to the central star.
In \k25~color, DUPLEX sources are spatially separated according
to the visibility of the central star:
DUPLEX sources with invisible central stars are redder than
SOLE sources while those with partially visible central stars
(\iras 08005$-$2356 and \iras 17423$-$1755) are bluer than 
SOLE sources.
This bifurcation among DUPLEX sources follows the split of
the SEDs into classes II and III: very high near-infrared excess 
in the class II DUPLEX sources makes their \k25~colors
bluer than the class III DUPLEX sources and even bluer than SOLE
sources.
The diagram thus not only signifies the difference between SOLE
and DUPLEX sources but also differentiates the partial/total 
obscuration of the central star in DUPLEX nebulae, and
should be a very useful tool in identifying the nature of
dust shell by near- to mid-infrared colors.

With new all-sky near-infrared surveys becoming available
(e.g. 2MASS and DENIS), near-infrared two-color diagrams
will also be a valuable tool in discriminating a particular
type of sources from a large data set.
Whitelock (1985) presented near-infrared ({\sl JHK}) photometry 
for 80 PNe and classified them into several types
in terms of the visibility of the central star due to dust 
obscuration and of the location in the near-infrared two-color 
\jh~vs. {\sl H--K} diagram.
Since the apparent morphological bifurcation is also partly based on 
the visibility of the central star, we adopt the classification
scheme of Whitelock (1985) and make use of such two-color diagrams.
Figure 10 is the \jh~vs. \hk~diagram with our sources that
have published near-infrared photometric data.
Also shown in the diagram are the regions according to
the object classification for PN candidates:
Nebula+Star, Nebula, Nebula+Dust, Star+Dust, and Miras
(\cite{whitelock85}).
Because of the dusty nature of PPNe, all of our sources are 
distributed over a linear diagonal region that corresponds to
the regions of Miras and Star+Dust objects.
There is a parallelism between the distributions of our targets 
and planetary nebulae in the diagram:
SOLE nebulae correspond to PNe with prominently 
visible central nuclei (Nebula+Star), whereas DUPLEX nebulae
correspond to dust enshrouded PNe (Nebula+Dust).
There also seems to be a border that separates the region of SOLE
sources from that of DUPLEX sources on the region of Miras
(dashed line in Fig. 10).
All SOLE nebulae are found on or 
blueward of the border and all DUPLEX nebulae are found 
on or redward of the border.
The fact that DUPLEX sources are redder than SOLE sources 
suggests that DUPLEX nebulae are associated 
with a larger amount of obscuring dusts than SOLE nebulae are
and this corroborates our hypothesis of the morphological bifurcation
being induced by the differing optical thickness in the two types 
of sources.

\subsubsection{The Inclination Angle Effect}
The inclination angle between the axis of symmetry with the line of 
sight can change the morphological appearance of an object.
We have seriously considered if the inclination angle effect
alone could explain all the observed differences between
SOLE and DUPLEX nebulae with questions such as if SOLE
nebulae are nearly pole-on DUPLEX nebulae or not.
In the following, we discuss that the evidence suggests otherwise.

Suppose that the dust shell structure of all existing PPNe are of
DUPLEX type, i.e., all dust shells have the same geometry of 
optically thick tori.
When sources are oriented edge-on ($90^{\circ}$ inclination 
angle), most of the star light is blocked by the dust torus
and we observe optical reflection nebulosities in the form of
more or less well-balanced bipolar lobes with a dust lane
(e.g. \iras 17150$-$3224 and \iras 17441$-$2411).
However, in other cases when the inclination is in some intermediary angle, 
sources should appear as imbalanced nebulae either with a dust lane 
(with larger intermediate inclination angle) or without a dust lane
(with smaller intermediate inclination angle), whose central stars
are seen off center in the nebulosities when visible.
The imbalance in the structure of nebulosities occurs because the far side of the
nebulosity (which is pointing away from us) will be at least
partially obscured by the near side of the optically thick dust torus,
as we see in the images of \iras 08005$-$2356 and \iras 17423$-$1755, for example.
This imbalance of brightness was also shown in simulated images of 
\iras 17441$-$2411 with 4 different inclination angles presented 
by Su et al.\ (1998, their Fig. 5).
According to their simulation, the central star is seen evidently
off-centered even at $30^{\circ}$ inclination angle.

As we have seen in our images of SOLE nebulae, 11 of 21 nebulae appear 
as well-balanced, smooth, and symmetric low surface brightness nebulosities 
with their central stars located at centers of the nebulae.
Following the discussion above, this is possible only when all 11
sources are oriented exactly pole-on or extremely close to pole-on.
Thus, if this is the case one has to explain why nearly half the 
objects are oriented at zero or near-zero inclination angles with
respect to us.
For instance, \iras 07134$+$1005 would be a prime example of a PPN viewed
nearly exactly pole-on because it shows an extended, 
almost circular reflection nebulosity of uniform brightness 
with its central star at the center of the nebulosity.
Nevertheless, the mid-infrared images of \iras 07134$+$1005 clearly
show a two-peaked, limb-brightened dust torus which suggests a 
non-zero inclination angle and radiative transfer calculations 
support a model of an 
equatorially enhanced dust torus viewed at an inclination angle
of $\sim 45^{\circ}$ (\cite{meixner97}).
If \iras 07134$+$1005 were really a DUPLEX source viewed at a
45$^{\circ}$ inclination angle, it should have appeared as either
an imbalanced nebula with the central star located off center
or a bipolar nebula with one of the lobes partially obscured.
However, that is not the case and thus the inclination angle 
effect can not simply explain the data.
The most reasonable interpretation of the data is that the 
optical thickness along the line of sight
is too low to cause any detectable difference, which is
also supported by radiation transfer calculations 
(edge-on $\tau_{9.7\um} \sim 0.03$, \cite{meixner97}).

As we have seen in the previous section, the distinction between
SOLE and DUPLEX sources is obvious and the subdivision among 
DUPLEX sources is remarkable in the \jk~vs. \k25 diagram (Fig.10).
One of the most peculiar aspects of the {\sl IRAS}/near-infrared
diagram is that the sources are not distributed linearly
as in the \jk~vs. \hk~diagram.
If the inclination angle effect were the main cause for the distinction
between SOLE and DUPLEX sources, the sources would have been 
distributed linearly with the region 
of DUPLEX sources with obscured central stars being located between 
the regions of SOLE sources and DUPLEX sources without central stars.
Alternatively, the absence of subdivision among SOLE sources 
corroborates our view of SOLE sources being associated with
optically thin dust tori: no matter what the inclination angle is
there is no partial obscuration in the SOLE sources and they
cluster as a single group in the \jk~vs. \k25 diagram.

It is of course possible that we come across a source whose orientation
is exactly or very close to pole-on.
In such cases, how the reflection nebulosities appear is not trivial.
No matter to which morphological type a source belongs, it is  
incredibly difficult to detect reflection nebulosity when the source 
is viewed pole-on
because the central star appears extremely prominent and the prominent
PSF spikes are likely to severely obscure the nebulosity.
The appearance of \iras 07430+1115, a SOLE PPN, does not fall into the 
morphological type of the DUPLEX nebulae, but the source still looks
different from other SOLE nebulae.
Although SED and two-color diagrams also suggest that this object 
is a SOLE source, it is possible that this is
a DUPLEX source oriented at or very close to pole-on.
However, we tentatively classify the source as a SOLE nebula and will 
not delve into the issue in this study.
The pole-on cases must be further investigated with at least
two-dimensional calculations.
Radiative transfer calculations with fully axisymmetric models 
would clarify these inclination angle effects with more certainty
and such calculations will be the focus of our future work.

\subsubsection{Non-Association with Nebulosities}
There are six sources we tentatively classify as stellar sources
because no reflection nebulosities are detected.
Neither PSF subtraction nor deconvolution achieves the perfect 
removal of the WFPC2 diffraction features, and thus,
there are always some residual diffraction artifacts
which can mask a low surface brightness reflection nebula.
In particular, our observations are insensitive to circular 
nebulosities with extremely low surface brightness, which can be
easily confused with the WFPC2 circular diffraction wings.

When viewed pole-on, SOLE sources would appear more or less like 
stellar sources due to a combined effect of the high star-to-nebula 
contrast and a confusion with the WFPC2 PSF artifact.
For example, \iras 04386+5722 and \iras 05113+1347 may be 
such cases.
The images of these sources barely show nebulosities and
the deconvolved images are almost nebula-free for \iras 04386+5722
whereas suggestive of a compact nebula for \iras 05113+1347.
However, their SEDs are of class IVa (Fig. 8) and their locations 
in two-color diagrams are within the region of SOLE sources
(an asterisk and a filled star in Figures 9 and 10).
The slight offset of \iras 04386+5722 towards blue in \k25~color 
in the \jk~vs. \k25 diagram can be explained considering the 
pole-on inclination angle effect: 
the source is of SOLE type and hence the stellar emission peak is 
present in the SED irrespective of the angle of inclination 
and hence its \jk~color would not be very different from other SOLE 
sources,
while its \k25~color can be bluer than other SOLE sources because
the stellar emission would be more prominent than other SOLE sources
with respect to the dust emission.
\iras 05113+1347 is located among other SOLE sources.

\iras 10158$-$2844 and \iras 15465+2818 have very bright 
central stars, but the amount of far-infrared excess is smaller
(Fig. 8).
The lack of far-infrared excess in these sources makes
the SED classification scheme of VHG89 inapplicable and 
also makes both have much bluer \k25~color than other sources
(\k25 = 3.48 and 3.12, respectively).
This is consistent with the evolutionary status of
these R Coronae Borealis stars, in which they
have had little recent mass loss but are associated
with very diffuse, extended circumstellar dust shells 
(\cite{gillett86}, \cite{waters89}).
However, if we consider the stellar component peaks alone
(which peak at $< 1\um$)
these objects will be found among DUPLEX sources (Fig. 10;
filled stars in the region of DUPLEX sources)
suggesting the presence of circumstellar dust rather close
to the central star, on-going mass loss, or pole-on 
inclination angle effects.
The nature of variability of \iras 10158$-$2844 is suspected
to be due to variable obscuration by circumstellar material
along the line of sight and the inclination angle of this
source is expected to be $\age 50^{\circ}$ (\cite{waelkens91}).
In fact, both of these sources show indications of a very recent 
mass loss (\cite{clayton97}, \cite{meixner99}).
Hence, a possible explanation for the non-detection of any 
nebulosities is that the column density of dust near the star in 
these sources is much lower than in SOLE sources.

\iras 20043+2653 is classified as an OH/IR source 
(cf. \cite{garcialario97}), and hence, the central star 
is likely be in the AGB or extreme AGB phase.
This interpretation fits well with the SED shape and
its very red \hk~color (both of which resemble to those 
of \iras 09452+1330, an extreme AGB star).
Therefore, the absence of any reflection nebulosity seems to 
suggest that \iras 20043+2653 has not yet developed one.
\iras 22142+5206 is classified as a young stellar
object embedded in a massive CO molecular cloud of 
$\sim 7300 \msun$ (\cite{dobashi98}).
Its SED shows significant far-infrared excess
without a stellar component (Fig. 8), which is a signature of
a class I young stellar object (e.g., \cite{wilking89}).
Its location in two-color diagrams is also consistent with 
the young stellar object interpretation.

\subsection{Origins of the PPN Morphological Bifurcation}
The origin of axisymmetry in many astrophysical systems is 
always of a great importance and there have been numerous
possible mechanisms for the creation of nebular morphologies 
(cf. \cite{livio97}). 
Instead of reviewing every possible means, we will focus
on the origins of the differing equatorial
density enhancement in the SOLE and DUPLEX nebulae.

\subsubsection{Galactic Height and Progenitor AGB Stellar Mass}
One of the strongest pieces of circumstantial evidence that is 
related to the morphological bifurcation is probably that bipolar
nebulae are preferentially found close to the plane of the Galaxy.
With a large sample of PNe, Corradi \& Schwartz (1995) 
found that bipolar PNe were distributed closer to the Galactic plane 
(scale height $z_{h} = 130$ pc with $|z| \ale 850$ pc) than elliptical PNe
($z_{h} = 325$ pc with $|z| \ale 1300$ pc).
Following this finding, they suggested that bipolar PNe have evolved 
from more massive progenitors than elliptical PNe, adopting
a lower limit of 1.5$\msun$ for the bipolar PN progenitors.
This correlation is also suggested by the galactic latitudes
and the $|z|$ values of our sources (Table 4).
DUPLEX sources, having a mean height of $220$ pc with a range of
$|z| \ale 520$, are more confined to the Galactic plane than
the SOLE sources, which have a mean height of $470$ pc with
a range of $|z| \ale 2100$.
Although a direct comparison between our values (mean Galactic heights) 
to the values obtained by Corradi \& Schwartz (1995; 
Galactic scale heights) is not possible, there certainly exists 
a parallelism in the ways these two types of PPNe and two types
of PNe are distributed in the Galaxy.
To test if the Galactic height distributions of the SOLE and DUPLEX 
sources are not exactly equal, we calculated the Kolmogorov--Smirnov (K--S)
statistic (0.417) and its significance level (0.186).
This means that there is $18.6 \%$ chance that the K--S statistic, 
the greatest difference between the two cumulative distribution 
functions of the Galactic heights, will be smaller than 0.417,
if both types of objects are from the same Galactic height
distribution.
Considering the fact that some SOLE sources do exist close to
the Galactic plane where DUPLEX sources are populated, the outcome 
of the K--S test is suggestive
that SOLE and DUPLEX sources are not distributed in the
exactly same manner.
Therefore, it is likely that more massive progenitor
AGB stars lead to DUPLEX PPNe and that bipolar PNe are the 
direct descendants of DUPLEX PPNe and elliptical PNe are the 
SOLE PPN offspring.
In fact, this is a very reasonable result in the context
of stellar evolution.
The majority of AGB stars become white dwarfs of roughly 0.6 
$\msun$ (\cite{schonberner81});
the more massive the progenitor, the more material 
the star has to dump into the circumstellar environment.
Therefore, PPNe from more massive progenitor AGB stars are
likely to have more obscuring material in the circumstellar
shell, which could completely block the central star
from the observer's view as in DUPLEX nebulae.
This is well demonstrated in Figure 10 as DUPLEX sources
being redder than SOLE sources.
Also consistent with the division of PPN into two classes is that
the degree of equatorial enhancement and the subsequent
evolution depend on the fundamental parameters of the
central star, especially, the stellar mass.

\subsubsection{Circumstellar Chemistry}
Because PPNe can be put into two groups in terms of photospheric
and/or circumstellar chemistry (C- or O-rich), 
we looked for a correlation between the circumstellar 
chemistry and morphological bifurcation.
When referring to the chemical type of a PPN, one needs 
to be cautious because a PPN can have both C- and O-rich
characteristics in the circumstellar environment above the
photosphere (e.g., \iras 08005$-$2356 and Red Rectangle).
The circumstellar chemistry can be determined mainly by the 
presence of a certain molecular species (e.g., OH in an O-rich
shell, \cite{hu94}; HCN in a C-rich shell, \cite{loup93}) 
or some infrared spectral feature (e.g., 9.7$\um$ feature in 
an O-rich shell; 21$\um$ feature in a C-rich shell, 
\cite{kwok89}).
On the other hand, the photospheric chemistry can be determined
by direct abundance measurements (e.g.,
\iras 02229+6208, \iras 07430+1115, \cite{reddy99b};
\iras 04296+3429, \cite{decin98};
\iras 05341+0852, \cite{reddy97};
\iras 06530$-$0213, \cite{reddy99};
\iras 18095+2704, \cite{klochkova95};
\iras 19114+0002, \cite{reddy99a};
\iras 20462+3416, \cite{garcialario97})
or the presence of optical photospheric features of C$_2$ or
C$_3$ molecules (e.g., \cite{hrivnak95}).
Comparison between the PPN morphology and photospheric/circumstellar chemistry
does not seem to yield any apparent correlation between the two.
Table 4 (column 7) summarizes the non-correlation between 
morphology and chemistry.

\subsubsection{Stellar Ages}
It may be possible to attribute the differing optical thickness
in SOLE and DUPLEX PPNe to their ages.
PPN dust shells expand with time and older shells tend to 
have smaller optical depth, and therefore, DUPLEX PPNe are 
expected to be younger than SOLE PPNe.
If one insists on a single evolutionary channel, DUPLEX PPNe
may be forerunners of SOLE PPNe.
However, bifurcating PPNe into SOLE and DUPLEX types by their 
ages does not appear to be possible because of the spectral 
types of these stars.
If DUPLEX PPNe were indeed younger than SOLE PPNe,
DUPLEX PPNe would have had the latest spectral types
possible (G to M types) and SOLE PPNe would have had
the earliest spectral types (A to F types).
However, SOLE PPNe include a number G types 
(e.g. \iras 04296$+$3429) and
DUPLEX PPNe include F types (e.g. \iras 08005$-$2356 and
Egg Nebula).
Given the horizontal evolutionary track in the HR diagram
during the PPN phase and its surpassingly short evolutionary 
time scale, it is not possible 
to conclude one type of PPNe is younger than the other.

\subsection{From the Dual PPN Morphology to the PN Morphology}
During the PPN phase, the two-layered PPN shell keeps 
expanding around the central post-AGB star while the surface 
temperature of the star continues to rise.
A fast wind initiates somewhere along the PPN phase and pushes 
the inner boundary of the PPN shell out to typical PN
dimensions, while shaping the boundary geometry and increasing 
the boundary density (\cite{kwok82}).
The central post-AGB star finally becomes hot enough to emit 
photoionizing photons, which illuminate the inner 
boundary of the circumstellar gas shell as a PN.
Because we observe dust-scattered light in PPNe and ionized gas 
emission in PNe, we can not trivially link the PPN and PN morphologies 
via a mere resemblance in the images.
We can, nevertheless, interpolate the PPN and PN morphologies
considering the circumstellar distribution of matter and see 
the PN morphological structures
(round, elliptical, and butterfly classes; \cite{balick87}) 
in the dual PPN morphology.
The bipolar shapes of DUPLEX PPNe 
(e.g., \iras 08005$-$2356, \iras 17150$-$3224) 
may be forerunners to the bipolar PNe 
(e.g., Hourglass Nebula, \cite{sahai95}).
The simply elongated SOLE nebulae 
(e.g., \iras 17436+5003, \iras 20462+3416)
may be precursors of elliptical PNe
(e.g., NGC 3132 (\cite{hst98}), IC 3568 (\cite{bond97})).
The multi-lobed SOLE nebulae
(e.g., \iras 22272+5435, \iras 06530$-$0213)
may be progenitors of complex PNe
(e.g., Stingray Nebula (\cite{bobrowsky98}), 
Cat's Eye Nebula (\cite{harrington95})).
We thus see that the development of the PPN axisymmetry in 
the superwind phase probably sets the stage for the 
emergence of axisymmetry in PNe.
Jet-like structures, however, are rather rare in PPNe
(e.g., \iras 17423$-$1755, which is a young PN) and 
this may suggest that the formation of jet-like structures
seen in PNe (e.g., NGC 5307, \cite{bond97}) does not share 
the same generating mechanism as the PPN axisymmetric 
structures.

\section{Conclusions}
After observing 27 PPN candidates with \hst,
we have found elongated low surface brightness reflection 
nebulosities around 21 sources.
We have also found 
that an optical reflection nebulosity can manifest itself in 
the form of a faint, elliptically elongated shell in addition to
the bipolar form.
The PPN circumstellar shell seems to be intrinsically 
axisymmetric (ellipticity $\sim$ 0.44) and we argue 
that the 
axisymmetry emerges in the superwind phase, the latter of the 
two-phased AGB mass loss epoch.
A morphological bifurcation exists among the PPN nebulosities:
of 21 extended nebulae, 11 are SOLE nebulae (e.g., \iras 07134+1005)
and 10 are DUPLEX nebulae (e.g., \iras 17150$-$3224).
We discuss how the morphological dichotomy is caused by 
the difference in optical thickness of the PPN circumstellar 
dust shells:
SOLE shells are optically thin whereas a DUPLEX shells
are optically thick.
The distinctness between SOLE and DUPLEX nebulae in terms of
optical thickness of the dust shells is evidenced by 
the correlation between the mid-infrared morphology of dust emission 
regions and optical morphology of reflection nebulosities,
the characteristic shapes of the SEDs, and the 
near- and {\sl IRAS}/near-infrared two-color diagrams.
We also discuss that the inclination alone may not be able to 
explain the well-balanced shape of reflection nebulosities with 
their central stars seen at the center.
Although we find no correlation between the circumstellar chemistry 
and morphology, we do find that DUPLEX sources tend to be found
closer to the Galactic plane than SOLE sources.
This suggests that 
DUPLEX PPNe probably originate from higher mass AGB progenitor 
stars than SOLE PPNe.
The origins of the apparent morphological bifurcation -- the 
equatorial density enhancement in the superwind -- remain 
inconclusive.
In addition to optical imaging of reflection nebulosities,
direct, high-resolution imaging of dust emission regions in 
mid-infrared wavelengths will be extremely important in studies 
of PPNe because optical
images of reflection nebulosities alone are not sufficient to
decipher the orientation of these objects. 
Similarly, future investigation will have to require at least 
two-dimensional, radiative transfer model calculations.

\acknowledgements
This research is based on observations with the NASA/ESA 
Hubble Space Telescope, obtained at the Space Telescope 
Science Institute (STScI), which is 
operated by the Association of Universities for Research in 
Astronomy, Inc. under NASA contract No. NAS5-26555.
Data reduction was done with routines provided in
the Space Telescope Science Data Analysis System (STSDAS), which is a 
software package for calibrating and analyzing data from \hst. 
STSDAS includes the same calibration routines as are used in the 
routine data processing pipeline, as well as general-purpose 
tools and enhancements to the Image Analysis and Reduction Facility 
(IRAF) distributed by the National Optical Astronomy 
Observatories, which are operated by the Association of 
Universities for Research in Astronomy, Inc., under cooperative 
agreement with the National Science Foundation.
This research also made use of the SIMBAD database, operated at 
CDS, Strasburg, France.
Ueta and Meixner are supported by NASA Grants GO-06737.01-95A 
and GO-06364.02-95A and NSF Career Award Grant AST 97-33697.
Bobrowsky is supported by STScI Grants GO-06364.01-94A and
GO-006737.02-95A.
We are thankful for the help that Chris Skinner gave us 
in setting up these observations.  
Chris J. Skinner, who died suddenly in October 1997, was a valued 
colleague and friend.   
We would also like to thank the STScI support staff for their 
guidance in setting up these observations and helping with 
the reduction of the data.

%Figure Captions
\newpage
\figcaption{(a)--(c): Images of SOLE nebulae
in the increasing order of their right ascension 
(north is up and east is left): 
the left-most frame shows the \iras ID and scale
of the object.
The tick marks show relative offsets in arcseconds.
The filter types are shown at the bottom of each frame
with ``+ RL'' indicating Richardson-Lucy deconvolution.
Wedges show the ranges of log-scaled flux density
to help readers visually illustrate the emission contrast.
See Table 1 for the star-to-nebula surface intensity 
ratio.}

\figcaption{(a)--(c): Images of DUPLEX nebulae.  
The displaying scheme follows Figure 1.  
See Table 2 for the star-to-nebula surface 
intensity ratio.}

\figcaption{Selected Images of stellar sources.
These images exemplify how PSF artifacts appear in
both reduced and deconvolved images.
The basic displaying scheme follows Figure 1.}

\figcaption{A schematic diagram illustrating the essential 
difference between SOLE and DUPLEX sources.  
[Top] In the SOLE sources, the superwind mass loss creates
marginally equatorially enhanced dust shell (gray area near 
the central star)
within a spherically symmetric the AGB wind shell.
Because the axisymmetric dust shell is {\it optically thin},
it permits star light to leak out in all directions (arrows), 
creating an elliptically elongated nebula (dashed line) 
with the bright central star.  
[Bottom] In the DUPLEX sources, the superwind mass creates
highly equatorially enhanced dust shell (black area near
the central star)
within a spherically symmetric the AGB wind shell.
Because the axisymmetric dust shell is {\it optically thick},
it permits starlight to leak out only along the 
biconical openings of the dust torus (arrows), creating
a bipolar, dumbbell-like nebulosity (dashed perimeter)
with the partially or completely obscured central star.
Relative sizes of components are not to scale.}

\figcaption{Composites of optical and mid-infrared images of four 
PPN candidates.
Mid-infrared contours are overlaid onto an optical grayscale image.
The top row shows examples of the SOLE-toroidal correlation 
(\iras 07134+1005: F410M and 11.8$\um$ and 
\iras 17436+5003: F410M and 12.5$\um$)
and the bottom row shows examples of the DUPLEX-core/elliptical 
correlation (\iras 17150$-$3224: F814W and 12.5$\um$ and 
\iras 16342$-$3814: F814W and 9.8$\um$).
Tickmarks show relative offsets from the center.}

\figcaption{Spectral energy distributions of SOLE sources in 
$\lambda F_{\lambda}$ (ergs s$^{-1}$ cm$^{-2}$) vs. 
$\lambda$ ($\um$).
The name of the source and its SED class assignment 
(\cite{veen89}; see text) are given in each frame. 
Photometric data are taken from our measurements and references listed 
in Table 4.}

\figcaption{Spectral energy distributions of DUPLEX sources in 
$\lambda F_{\lambda}$ (ergs s$^{-1}$ cm$^{-2}$) vs. 
$\lambda$ ($\um$).
Conventions follow those of Fig. 6.}

\figcaption{Spectral energy distributions of stellar sources in 
$\lambda F_{\lambda}$ (ergs s$^{-1}$ cm$^{-2}$) vs. 
$\lambda$ ($\um$).
Conventions follow those of Fig. 6.}

\figcaption{%
{\sl IRAS}/Near-infrared two-color (\jk~vs. \k25) diagram for our sources.
Circles, triangles, and stars are respectively SOLE, DUPLEX, and 
stellar sources.
The asterisk is \iras 04386+5722, a stellar source whose 
chemistry is undetermined.
The numbers associated with symbols refer to the source numbers
given in Table 4 (column 1).
Two stellar sources, \iras 10158$-$2844 and \iras 15465+2818,
and one DUPLEX source, \iras 09452+1330,
are located off the diagram: (\k25, {\sl J--K}) = (3.48, 1.59), 
(3.12, 2.11), and (10.21, 6.54), respectively.
This diagram displays the color of the star vs. the ratio
of the dust peak to the stellar peak in the SED.
Sources are more clearly clustered into three groups: SOLE 
and DUPLEX sources with a
totally obscured central star (w/o Star), and DUPLEX sources
with a visible central star (w/ Star). 
Each subdivision is labeled in the diagram.
The dashed line indicates a fiducial division between SOLE and 
DUPLEX sources.
All sources above the line are of DUPLEX type and below the 
line are of SOLE type.}

\figcaption{%
Near-infrared two-color (\jh~vs. {\sl H--K}) diagram for our sources.
Conventions follow those of Fig. 9.
Two AGB stars, \iras 09452+1330 (C-rich) and \iras 20043+2653
(O-rich) are located off the diagram: 
({\sl H--K\/}, {\sl J--H}) = (3.03, 3.51) 
and (2.79, 0.30), respectively.
Whitelock's classification of planetary nebula (1985) is also 
shown in the diagram: the regions of Nebula+Star, Nebula+Dust, 
Star+Dust, Nebula, and Miras.
Sources are clustered diagonally in the diagram, which 
indicates the black-body temperature decreasing to the upper right.
The dashed line indicates a fiducial division between SOLE and 
DUPLEX sources.
All sources on the right of the line are of DUPLEX type and on 
the left of the line are of SOLE type.}

\end{document}